\documentclass[%
 reprint,
 twocolumn,
 amsmath,amssymb,
 aps,
 prl,
 nofootinbib,
 tightenlines,
 nobibnotes
]{revtex4-2}

\usepackage{graphicx}
\usepackage{dcolumn}
\usepackage{bm}
\usepackage{hyperref}

\usepackage{color}

\usepackage{gensymb}

\definecolor{Green}{RGB}{0,100,0}
\definecolor{Purple}{RGB}{102,0,255}
\definecolor{Blue}{RGB}{51,153,255}
\definecolor{Red}{RGB}{151,010,010}

\definecolor{Orange}{RGB}{255,69,0}

\begin{document}

\title{Hydrodynamics Explanation for the Splitting of Higher-charge Optical Vortices}

\author{Andrew A. Voitiv$^1$, Jasmine M. Andersen$^1$, Patrick C. Ford$^1$, Mark T. Lusk$^2$,}
\author{Mark E. Siemens$^1$}%
\email[]{msiemens@du.edu}%
\affiliation{$^1$Department of Physics and Astronomy, University of Denver, 2112 E. Wesley Avenue, Denver, CO 80208, USA
}%
\affiliation{
$^2$Department of Physics, Colorado School of Mines, 1500 Illinois Street, Golden, CO 80401, USA
}%

\date{\today}

\begin{abstract}
We show that a two-dimensional hydrodynamics model provides
a physical explanation for the splitting of higher-charge optical vortices under elliptical deformations. The model is applicable to laser light and quantum fluids alike. The study delineates vortex breakups from vortex unions under different forms of asymmetry in the beam, and it is also applied to explain the motion of intact higher-charge vortices.
\end{abstract}

\maketitle

It is well-known that optical vortices, characterized by swirling phase gradients that wrap over $2\pi \ell$ for topological charge $\ell$, are topologically protected, but for higher-charge vortices the more relevant concept is charge conservation. It is widely documented that higher-charge vortices are prone to decomposition: an $\ell$-charged vortex splits into $|\ell|$ unit-charge vortices unless a pure vortex mode is propagated in a spatially-uniform medium \cite{Freund1993OpticalMedia}.  Evidence of vortex splitting is attributed to low coherent background light \cite{Basistiy1993OpticsDislocations}, imperfections with gratings or finite-pixelated spatial light-modulators \cite{Basistiy1993OpticsDislocations,Kumar2013StabilityModulators}, elliptical perturbations of the laser beam itself \cite{Dennis2006RowsBeam}, or other critical point and saddle perturbations in the phase of the beam (such as those that arise from the presence of an opposite-charge vortex) \cite{Freund1999CriticalFields, Freund1995SaddlesFields, Berry2001KnottedWaves}. Vortex stability has been of interest in linear and nonlinear singular optics applications \cite{Liang2020SplittingBeams, Maleev2003CompositeVortices, Gan2009StabilizationNonlinearity, Kumar2011CraftingVortices, Kumar2013StabilityModulators, Ricci2012InstabilityInterferometer}, and an understanding of the physical origin of vortex splitting is important because it could enable enhanced engineering and control of optical vortex applications \cite{Shen2019OpticalSingularities}, such as communications through atmosphere \cite{Gbur2008VortexConservation}, or it could provide new insights 
into the physics of vortices in other systems such as quantum fluids \cite{Rokhsar1997VortexGases,Nilsen2006VelocityCondensates, Groszek2018MotionCondensates, Zhu2021DynamicsFluid}.

\begin{figure}[h!]
\centering
\includegraphics[width=\linewidth]{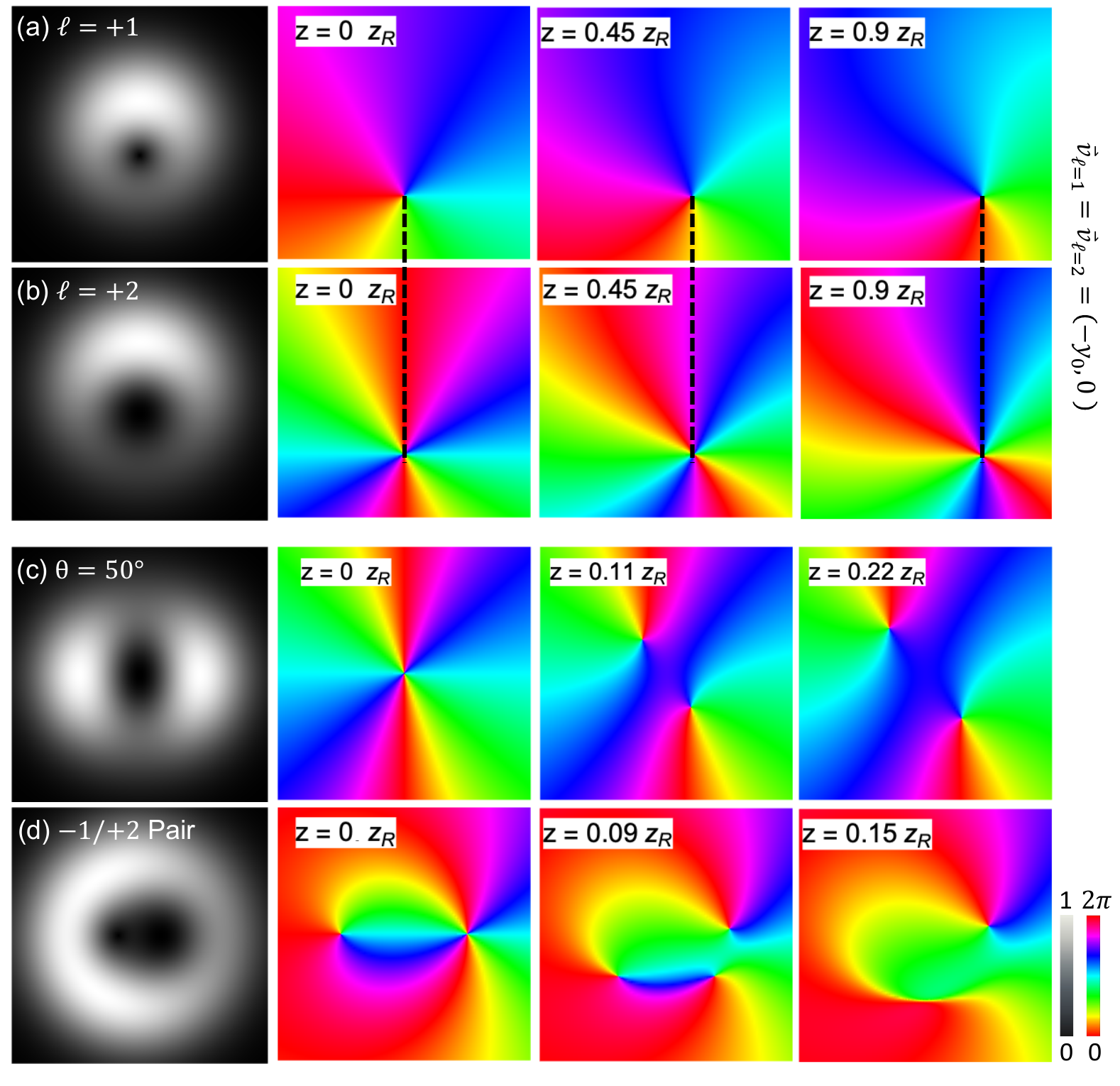}
\caption{(a) Off-set vortex in a Gaussian laser beam moves at a constant rate. (b) Higher-charge vortex at the same off-set moves at the same constant rate. (c) An elliptically deformed higher-charge vortex splits into unit vortices. (d) A higher-charge vortex (right-shifted) splits in the presence of an opposite-charge vortex (left-shifted), with subsequent annihilation events between $\pm 1$ pairs. Propagation is in terms of the Rayleigh length, $z_{R}$.}
\label{fig:fourexamples}
\end{figure}

To motivate the need for a physical understanding of this topic, consider a model of the propagation of a higher-charge vortex off-center in a Gaussian laser beam in Fig. \ref{fig:fourexamples} (b). A similar setting was studied in the context of vortex dynamics in a 2D BEC \cite{Groszek2018MotionCondensates}, where the velocity of a vortex was predicted to move at a rate that scaled with its charge. Due to atomic interactions in the system, the vortex in that study split into unit charges \cite{Rokhsar1997VortexGases, Leanhardt2002ImprintingPhases}; but in a small window of time, the authors measured numerical results between a unit-charged velocity and the predicted charge-scaled velocity. Those measurements approached the authors' charge-scaled velocity prediction using a disk trap (rather than harmonic) for the nonlinear sample. In a purely linear system with no trap, and with ``linear core'' unit-charge vortices, we find that a higher-charged vortex moves not with a velocity that scales with its charge, but rather it moves at the same rate as a unit-charge vortex, as shown in Fig. \ref{fig:fourexamples} (b). Further, in this linear optics model, there is no splitting of the off-center vortex, which is consistent with work showing that arbitrary distributions of vortices preserve their patterns from near- to far-field  \cite{Indebetouw1993OpticalPropagation}.

In this Letter, we provide a physical explanation for higher-charge vortex motion and splitting in linear optics based on two-dimensional, compressible hydrodynamics. We find that circular higher-charge vortices move as a collection of independent unit-charge vortices, and apparent vortex ``splitting'' arises from the presence of elliptical deformations of the vortices. This study contributes to the explanations of vortex splitting in the singular optics works cited above with new insights on vortex dynamics and so-called ``interactions'', with findings that are relevant to vortices in 2D quantum fluids, in optics for communications, and in optical speckle.

We begin by describing the hydrodynamic model. It has been shown that this model comprehensively accounts for the evolution of unit-charge vortices as functions of changes in the host background field \cite{Andersen2021HydrodynamicsFluids}, involving any configuration of vortices of mixed signs of $\pm 1$ or unit-charge vortices that are initially elliptical. In the model, the total field containing any number of vortices embedded in a host beam is separated into two products: $\psi_{\mathrm{v}}$ for \textit{one} vortex under consideration, and $\psi_{\mathrm{bg}}$ for the \textit{rest} of the field that comprises the ``background,'' together composing the total field $\psi = \psi_{\mathrm{v}} \, \psi_{\mathrm{bg}}$. Analysis of the paraxial equation under a Madelung (hydrodynamic) transformation shows that the velocity of the vortex at propagation step $z$ is \cite{Andersen2021HydrodynamicsFluids}
\begin{equation} \label{velocity}
    \vec{v}_{\mathrm{v}} = \vec{\nabla}_{\perp} \varphi_{\mathrm{bg}} - \frac{\mathbf{V}^2_{\mathrm{v}}}{J_{\mathrm{v}}} \boldsymbol{\sigma}_0 \vec{\nabla}_{\perp} \ln{{\rho_{\mathrm{bg}}}},
\end{equation}
which is evaluated at the $(x_0,y_0)$ location of the vortex, and where $\varphi_{\mathrm{bg}}$ and $\rho_{\mathrm{bg}}$ are the phase and amplitude of the background field, respectively. The other terms, $\mathbf{V}_{\mathrm{v}}$, $J_{\mathrm{v}}$, and $\boldsymbol{\sigma}_0$, are determined by the deformation of the target vortex, $\psi_{\mathrm{v}}$, and are defined as:
\begin{equation}
    J_{\mathrm{v}} = \sec{\theta}, \hspace{.1in} \boldsymbol{\sigma}_0 = 
        \begin{pmatrix}
        0 & -1 \\
        1 & 0
        \end{pmatrix},
\end{equation}
\begin{equation*} \label{modifier}
    \mathbf{V}_{\mathrm{v}} = 
        \begin{pmatrix}
        \cos{\xi} & -\sin{\xi} \\
        \sin{\xi} & \cos{\xi}
        \end{pmatrix}
        \begin{pmatrix}
        1 & 0 \\
        0 & \sec{\theta}
        \end{pmatrix}
        \begin{pmatrix}
        \cos{\xi} & -\sin{\xi} \\
        \sin{\xi} & \cos{\xi}
        \end{pmatrix}^{-1}.
\end{equation*}
The deformation yields the vortex ellipticity, which can be described by ``virtually tilting'' the vortex along two angles with respect to the propagation axis, $z$. These angles are the polar lean, $\theta$, and azimuthal orientation, $\xi$, which are analogous to the angles on a modal Poincar\'e Sphere \cite{Gutierrez-Cuevas2019GeneralizedVectors}. The polar lean, $\theta$, stretches the vortex (causing ellipticity) while the azimuthal angle, $\xi$, rotates it about its centroid.

The velocity in Eqn. \ref{velocity} determines the subsequent kinematics, since the vortex position $\vec r_\mathrm{v}$ at propagated position $z_0+\Delta z$ can be determined by $\vec r_{\mathrm{v}}(z_0+\Delta z) = \vec{r}_{\mathrm{v}}(z_0)+ \vec{v}_\mathrm{v} \Delta z$.  Eq. \ref{velocity} therefore provides a \textit{physical explanation} for observed vortex motion in terms of hydrodynamics---namely the gradients of the background field phase and amplitude, as well as the virtual tilt of the target vortex. For a single $\ell = +1$ vortex in a Gaussian beam, Eqn. \ref{velocity} was shown to yield $\vec{v}_{\mathrm{v}} = (-y, x)$ \cite{Andersen2019CharacterizingHolography}, which is then evaluated at the location of the vortex, $(x_0, y_0)$, thus predicting a stationary, centered vortex for unperturbed Laguerre-Gaussian beams. An $\ell=+1$ vortex moves with a velocity of $v_x = -y_0$ \cite{Rozas1997PropagationVortices, Andersen2021HydrodynamicsFluids}, as shown in Fig. \ref{fig:fourexamples} (a) with $y_0=-0.5$.

We now turn to a higher-charge vortex in a Gaussian, whose initial condition is typically written as
\begin{equation*} \label{typical}
    \psi|_{\small z=0} = \sqrt{\frac{2}{\pi \, |\ell|!}} \, e^{\frac{-(x^2+y^2)}{2}} \left[ (x - x_0) + i \, \mathrm{Sgn}[\ell] (y - y_0) \right]^{|\ell|},
\end{equation*}
where the paraxial equation has been non-dimensionalized using $x \rightarrow x \, w_0/\sqrt{2}$, $y \rightarrow y \, w_0/\sqrt{2}$, and $z \rightarrow z \, k \, w_0^2 /2$ for wave number $k$ and beam waist $w_0$. To include the possibility of vortex deformation, we write
\begin{multline} \label{multi}
    \psi|_{\small z=0}= \sqrt{\frac{2}{\pi \, |\ell|!}} \, e^{\frac{-(x^2+y^2)}{2}} \biggl[ \left[ (x - x_0) + i \, (y - y_0) \cos{\theta} \right] \cos{\xi} \\ 
    + \left[ (y - y_0) - i \, (x - x_0) \cos{\theta} \right] \sin{\xi} \biggr]^{|\ell|}.
\end{multline}
The initial condition of Eqn. \ref{multi} can be propagated by Fresnel integration of the paraxial field \cite{Andersen2021HydrodynamicsFluids}, and vortex tracking (finding the intersections of real and imaginary zero-contours) confirms that a higher-charge, circular vortex does not split and has the same velocity as a unit-charge vortex---as depicted in the first two rows of Fig. \ref{fig:fourexamples}. Eqn. \ref{multi} is not the only way to express the initial condition for an (elliptical) vortex; other forms are possible \cite{Indebetouw1993OpticalPropagation}, which may introduce changes in the background field that will affect the subsequent vortex motion.

The observed motion of Fig. \ref{fig:fourexamples} (b) is not consistent with earlier work indicating that the vortex velocity scales linearly with $\ell$ \cite{Groszek2018MotionCondensates}. However, it is consistent with a direct application of the vortex kinetic relation of Eqn. \ref{velocity} under the proviso that the charge-$\ell$ vortex be viewed as a set of identical, coincident unit-charge vortices. For the $\ell=2$ case, the background field of one of the constituent unit-charge vortices is the product of a Gaussian beam and its clone: $\psi_{\mathrm{bg}} = \sqrt{\frac{2}{\pi}} \, e^{\frac{-(x^2 + y^2)}{2}} ((x-x_0) + i \, (y-y_0))$. The evaluation of Equation \ref{velocity} then yields $\vec{v}_{\mathrm{v}} = (-y, x)$, the same velocity as the unit-charge case. This result can be summarized as indicating that a circular charge-$\ell$ vortex is equivalent to $\ell$ identical, unit-charge vortices that do not interact~\cite{Rozas1997PropagationVortices, Roux2004CouplingVortices, Andersen2021HydrodynamicsFluids}.

\begin{figure}[h!]
\centering
\includegraphics[width=0.9\linewidth]{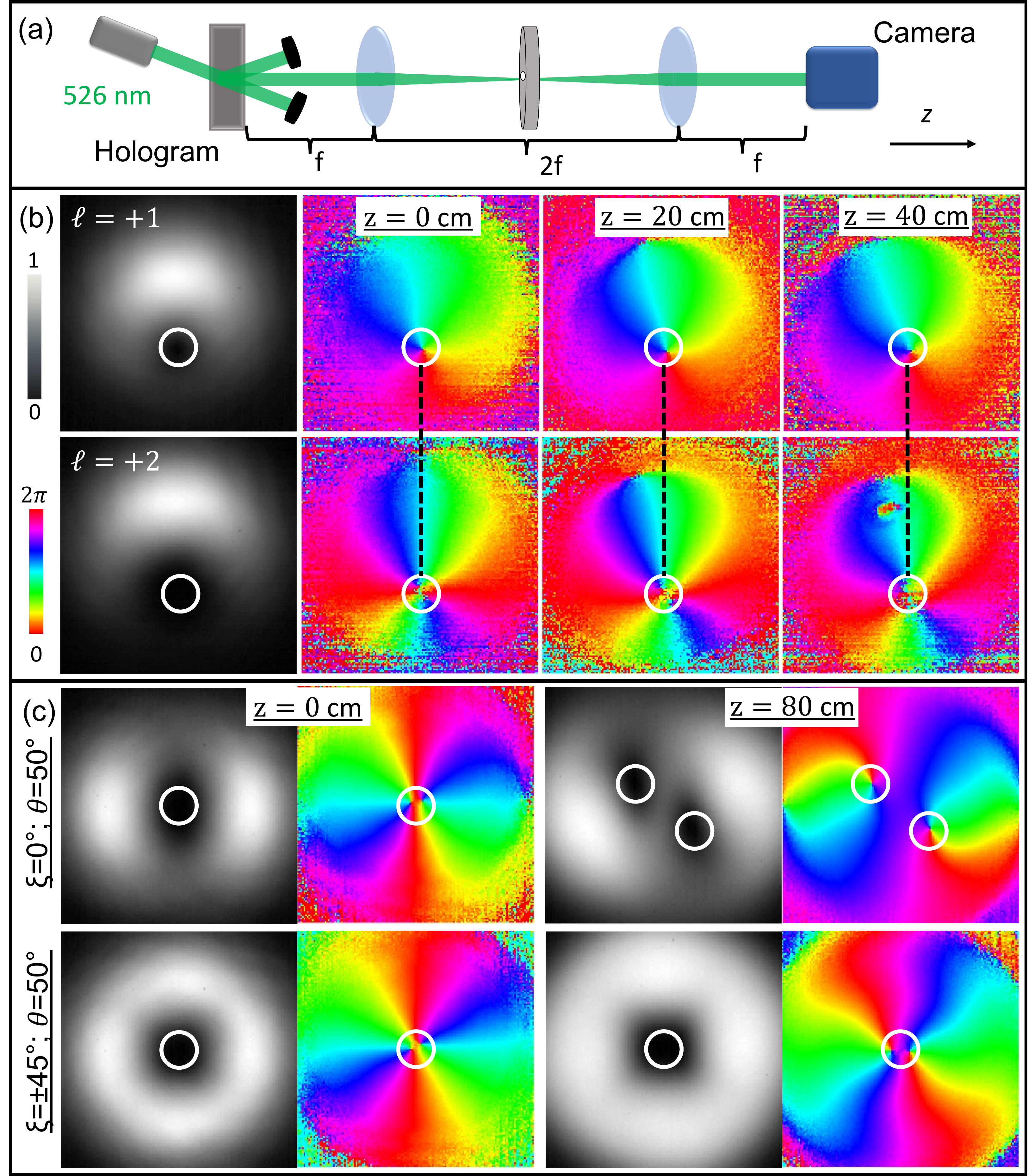}
\caption{(a) Simplified experimental schematic to measure vortex dynamics. A 
collimated Gaussian beam (wavelength $\lambda = 526$ nm) is transmitted through a hologram to construct the modes studied, with generated beam waist of $w_0 = 0.5$ mm. The beam is 4f-imaged from its generation to a detector to measure the transverse irradiance of the beam at increasing propagation steps. (b) Experimental counterpart to Fig. \ref{fig:fourexamples} (a) and (b). White circles, all of the same size, are drawn to highlight the vortices in the phase, $\mathrm{Arg}[\psi]$, measured via phase-shifting digital holography \cite{Andersen2019CharacterizingHolography}. Vortex separation is evident in the phase maps due to experimental limitations \cite{Kumar2013StabilityModulators}, but this separation is of equal scale whether the vortex is centered or off-center in the absence of tilt-induced splitting.  (c) Experimental verification of tilted-vortex splitting (top row) and (bottom row) a case where vortex interactions are cancelled \cite{Roux2004CouplingVortices} to not split. Each plot is about $1.5 w_0$ in size.}
\label{fig:experiment}
\end{figure}

Surprisingly, this result shows that a higher-charge vortex may remain intact despite being offset from the center of the beam. In practice, any tiny experimental imperfection in the vortex generation will lead to a separation of the vortices. This is even evident in numerical models. For example, if the mode is propagated using a Laguerre-Gaussian modal decomposition with limited accuracy, the finite modal content and approximated coefficients result in vortex separation (even if slight). The Fresnel-integrated models shown here do not have this limitation, but any experimental implementation will. Fig. \ref{fig:experiment} (a) shows experimental results that have this ``imperfect mode'' vortex separation but still demonstrate equal rates of motion between a $+1$ and $+2$ vortex.

\begin{figure}[h!]
\centering
\includegraphics[width=0.9\linewidth]{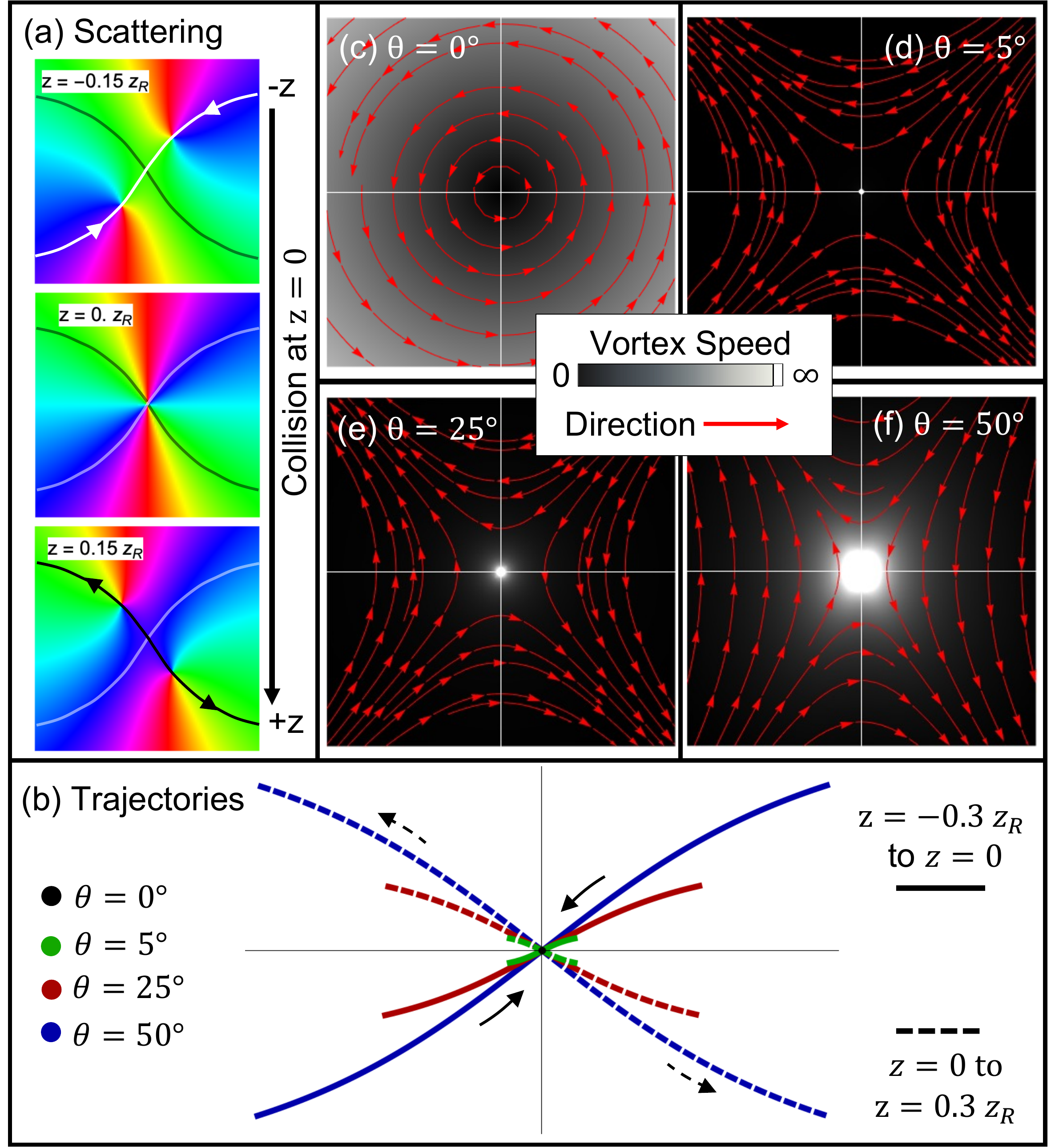}
\caption{(a) Model field of a tilted $(\theta=50\degree)$ vortex, propagated and back-propagated revealing a ``collision'' at the initial condition at $z=0$. (b) Analytical, two-dimensional trajectories of the same model field, as a function of propagation $(z)$, for different initial deformations $(\theta)$. The circular vortex, black data, remains stationary at the origin. (c)-(f) Combined stream-line plots (red) and density plots (grayscale) showing both the direction and speed, respectively, of vortex B at given infinitesimal displacement $(\delta x, \delta y)$ from the origin. Plotted over $-0.01w_0 \leq \delta x \leq 0.01w_0$ and $-0.01w_0 \leq \delta y \leq 0.01w_0$.}
\label{fig:vectors}
\end{figure}

Now consider a non-circular vortex characterized by non-zero tilt angles in Eqn. \ref{multi} and with a total charge of $\ell=2$. For mathematical simplification (but with no loss in generality towards the tendency to split), place this vortex at the center of the Gaussian beam and take $\xi = 0 \degree$. Once again, Fresnel integration can be used to obtain the field and the position of singularities as a function of time. These can be back-propagated as well, revealing that the tilted $\ell=2$ vortex at $z=0$ is simply a special case of two unit-charge vortices that happen to be co-located at that instant in their propagation. As shown in Fig. \ref{fig:vectors} (a), the dynamics are those of two tilted, single-charge vortices that collide and subsequently scatter off one another at $z=0$. Analytical trajectories of  the individual unit-charge vortices, found by real and imaginary zero-contour crossings of the field, are shown in Fig. \ref{fig:vectors} (b) for different initial deformations. We focus on the moment of impact at $z=0$ and carry out a hydrodynamic analysis to explain why and how an elliptical higher-charge vortex appears to split into unit-charge constituents that subsequently follow these plotted trajectories.

Motivated by our decomposition of circular charge-two vortices, we view the initial state as a special case in which two identically tilted single-charge vortices happen to be initially co-located at the origin. A more general initial condition has one of the vortices (A) at the origin, and the other (B) at position $(\delta x_B, \delta y_B)$, arbitrarily close to the origin. As before, and provided B is not at the origin, its velocity can be obtained using the kinetic relation of Eqn. \ref{velocity}. (The evaluation of Eqn. \ref{velocity} is undefined if vortex B is at the origin, coincident with A.) The associated background field is:
\begin{multline*}
    \psi_{\mathrm{bg}}|_{\small z=0} = \sqrt{\frac{2}{\pi}} e^{\frac{-(x^2+y^2)}{2}} \, \biggl[ \left[ x + i \, y \cos{\theta} \right] \cos{\xi} \\ 
    + \left[ y - i \, x \cos{\theta} \right] \sin{\xi} \biggr].
\end{multline*}
Using this, the resultant calculation of Eqn. \ref{velocity} yields the following initial velocity of B:
\begin{multline} \label{tiltvelo}
    \vec{v}_{\mathrm{B}}|_{\small z=0}\\
    = \frac{1}{x_B^2 + y_B^2 \cos^2{\theta}}
    \begin{pmatrix}
    - y_B \cos{\theta} \left[ 1 + x_B^2 + (y_B^2 - 1) \cos^2{\theta} \right] \\
    x_B \left[ x_B^2 - 1 + (1 + y_B^2) \cos^2{\theta} \right] \sec{\theta}
    \end{pmatrix}.
\end{multline}
This expression can be visualized, as shown in Fig. \ref{fig:vectors} (c)-(f), as a continuum of vortex B velocities, with the velocity at any given point $(\delta x, \delta y)$ associated vortex B's location at $z=0$. For the special case of no polar tilt, panel (c), we have previously found that the velocity of B is well-defined at the origin---here the zero-speed is depicted by black in the center. However, if there is \textit{any} tilt ($\theta$ $>0\degree$), then the $+2$ vortex splits as shown in Figs. \ref{fig:fourexamples} (c), \ref{fig:experiment} (b), and \ref{fig:vectors} (a) for a chosen value of $\theta = 50 \degree$. Panels (d)-(f), in contrast to panel (c), reveal a divergent speed, and undefined velocity, at the origin (over-saturated white) with corresponding stream-lines (red) which indicate that vortex B will flow away (``scatter'') from the center for any infinitesimal location $(\delta x, \delta y)$ from vortex A. As in the previous examples, there is nothing unique about the case of $\ell=+2$: we find that any higher-charge ($|\ell| > 1$) tilted vortex splits into $|\ell|$ unit, tilted vortices with propagation \cite{Liang2020SplittingBeams}.

The interpretation of vortex splitting as two unit-vortices scattering at $z=0$ has a flavor of ``vortex interactions'' \cite{Rozas1997PropagationVortices, Roux2004CouplingVortices, Andersen2021HydrodynamicsFluids}, and indeed many have already studied how deformed (or elliptical or ``noncannonical'') vortices couple with each other in linear optics. In particular,
Roux considered the case of a tilted $\ell=+2$ vortex and quantitatively predicted that if the two vortices had (an analogous) azimuthal tilt difference of $\xi = 90 \degree$, there would be no splitting---no interactions, no ``scattering''---even in the presence of non-zero $\theta$ \cite{Roux2004CouplingVortices}. From a hydrodynamics standpoint, this can be easily verified with the machinery above by simply repeating the steps without the simplification of $\xi = 0\degree$. Many more interesting dynamics are possible with arbitrary $\xi$ (so too with arbitrary numbers of vortices). 

We provide experimental confirmation of the prediction of Roux in Fig. \ref{fig:experiment}, along with experimental confirmation of the same rates of motion between an $\ell=+1$ and $+2$ vortex (taken from a setup identical to that in \cite{Andersen2021HydrodynamicsFluids}). We partly include this data to emphasize that even though no vortex-vortex interactions are predicted or engineered, there is still vortex separation due to experimental limitations of finite spatial light modulator pixel sizes or any slight perturbations to the host laser beam itself \cite{Kumar2013StabilityModulators, Dennis2006RowsBeam, Liang2020SplittingBeams}. Further, the inherent vortex separation shows that the small-displacements, $(\delta x, \delta y)$, of vortex B above are physically justified, as the vortices still scatter on the same predicted trajectories as if they had been purely coincident, as in Fig. \ref{fig:fourexamples} (c). Lastly, this example shows that different possible vortex configurations can be engineered towards different applications using the main features of the hydrodynamics model.

Finally, we do not want to leave the impression that an undeformed higher-order vortex provides a path toward non-splitting vortices in general, so we consider a higher-charge vortex in the presence of an oppositely-charged vortex (similar results are found if this additional vortex has any tilt other than that of the higher-charge vortex) \cite{Freund1995SaddlesFields}. Consider an initial condition that has a host Gaussian with a $-1$ vortex off-centered to the left and a $+2$ vortex off-centered to the right:
\begin{equation} \label{onetwo}
    \psi|_{\small z=0} \propto e^{\frac{-(x^2+y^2)}{2}} \left[ (x + x_0) - i \, y \right] \times \left[ (x-x_0) + i \, y \right]^{2}.
\end{equation}
The vortices in this field show behavior of ``splitting'' of the $+2$ vortex on the right, and subsequent annihilation between one of the $+1$ vortices with the leftward $-1$ vortex. The motion of the vortices in this system is shown in Fig. \ref{fig:fourexamples} (d).

Next, apply the hydrodynamic model to predict the observed splitting of the $\ell=+2$ vortex caused by the presence of the -1 vortex. Take the perspective of targeting one of the $+1$ vortices that make up the $+2$ vortex, such that the background field is
\begin{equation*}
    \psi_{\mathrm{bg}}|_{\small z=0} \propto e^{\frac{-(x^2+y^2)}{2}} \left[ (x + x_0) - i \, y \right] \times \left[ (x-x_0) + i \, y \right].
\end{equation*}
As all vortices are initially circular by construction, evaluating Eqn. \ref{velocity} for both of the $+1$ vortices that make up the $+2$ vortex yields a rather interesting initial velocity that is the same for each of them: $\vec{v}_{x}|_{\small z=0} = 0$ and $\vec{v}_{y}|_{\small z=0} = x_0^{-1} + x_0$ (we remind the reader of the non-dimensionalized setting). This is the same initial velocity result as predicted for a pair of $\pm 1$ vortices \cite{Andersen2021HydrodynamicsFluids} which annihilate after propagation in a semi-circular arc. The result is unsurprising, in fact, because we have previously established that the two circular $+1$ vortices do not interact with each other, but they are each affected by the $-1$ vortex.

It was shown that, for a typical $\pm 1$ vortex pair which annihilate, the ``virtual tilt'' $\theta$ evolves \textit{immediately} from the initial condition \cite{Andersen2021HydrodynamicsFluids}. Thus, we need only propagate the field an infinitesimal amount and then invoke the entire tilt-dependent expression of Eqn. \ref{velocity}. Evaluating the velocity under those conditions shows, again, that the velocity is undefined at the composite $\ell=2$ vortex location. However, it is defined at arbitrarily small displacements, $(\delta x, \delta y)$, from this location. Once separated, the velocities of each $+1$ vortex are well-defined and determined at each propagation step, for this particular example or any arbitrary arrangement of vortices.

In summary, we have shown that a hydrodynamic model of compressible, two-dimensional fluids provides a predictive physical explanation for optical vortex splitting. This splitting has been referred to as a ``perturbation'' of the higher-order, deformed vortex, but our calculations show a singularity in the equation of motion for the vortex at one instant in time/propagation. From this point, unit-charge vortices will follow the possible allowed contours (given by the hydrodynamic velocity) out of the undefined location. This behavior can be viewed as a scattering event among vortices as they overlap for a brief moment in time/propagation on their respective trajectories. Lastly, the model used here is equally applicable to quantum gases \cite{Zhu2021DynamicsFluid}, where application of the same hydrodynamic equations will naturally incorporate the influence of nonlinearities on the dynamics of higher-order charges.

\noindent \textbf{Funding.} W. M. Keck Foundation; NSF (1553905)

\noindent \textbf{Disclosures.} The authors declare no conflicts of interest.

\noindent \textbf{Data availability.} Data underlying all results presented are available from the authors upon reasonable request.

\bibliography{Refs}

\end{document}